\documentclass[12pt]{article}

\usepackage{amsmath,amssymb}
\usepackage{graphicx}

\makeatletter

\renewcommand\section{\@startsection{section}{1}{\z@}
                                   {-3.5ex \@plus -1ex \@minus -.2ex}
                                   {2.3ex \@plus .2ex}
                                   {\normalfont\large\bfseries}}
\renewcommand\subsection{\@startsection{subsection}{2}{\z@}
                                   {-3.25ex\@plus -1ex \@minus -.2ex}
                                   {1.5ex \@plus .2ex}
                                   {\normalfont\normalsize\bfseries}}
\renewcommand\subsubsection{\@startsection{subsubsection}{3}{\z@}
                                   {-3.25ex\@plus -1ex \@minus -.2ex}
                                   {1.5ex \@plus .2ex}
                                   {\normalfont\normalsize\bfseries}}
\renewcommand\paragraph{\@startsection{paragraph}{4}{\z@}
                                   {3.25ex \@plus1ex \@minus.2ex}
                                   {-1em}
                                   {\normalfont\normalsize\bfseries}}

\makeatother

\newcommand{\be}{\begin{equation}}
\newcommand{\ee}{\end{equation}}
\newcommand{\bea}{\begin{eqnarray}}
\newcommand{\eea}{\end{eqnarray}}
\newcommand{\ba}{\begin{array}}
\newcommand{\ea}{\end{array}}

\newcommand{\id}{\hbox{1\kern-.27em l}}
\newcommand{\ZZ}{\mathbb{Z}}

\newcommand{\half}{ {\textstyle \frac{1}{2}  } }

\newcommand{\ga}{\gamma}

\newcommand{\si}{\sigma}

\newcommand{\om}{\omega}

\newcommand{\tha}{\theta}

\newcommand{\cN}{\mathcal{N}}

\newcommand{\rar}{\rightarrow}
\newcommand{\non}{\nonumber}

\newcommand{\SU}{\mathrm{SU}}
\newcommand{\SO}{\mathrm{SO}}
\newcommand{\SL}{\mathrm{SL}}
\newcommand{\Sp}{\mathrm{Sp}}
\newcommand{\su}{\mathrm{su}}
\newcommand{\so}{\mathrm{so}}
\newcommand{\spl}{\mathrm{sp}}
\newcommand{\Spin}{\mathrm{Spin}}
\newcommand{\Pin}{\mathrm{Pin}}
\newcommand{\U}{\mathrm{U}}
\newcommand{\ul}{\mathrm{u}}

\newcommand{\ts}{\textstyle}

\newcommand{\beq}{\begin{equation}}
\newcommand{\eeq}{\end{equation}}

\begin{document}

\begin{center}

\vspace*{5mm}
{\Large\sf   Low-energy spectrum of $\cN = 4$ super-Yang-Mills on $T^3$\,: \\ 
flat connections, bound states at threshold, and $S$-duality}

\vspace*{5mm}
{\large M{\aa}ns Henningson, Niclas Wyllard}

\vspace*{5mm}
Department of Fundamental Physics\\
Chalmers University of Technology\\
S-412 96 G\"oteborg, Sweden\\[3mm]
{\tt mans,wyllard@fy.chalmers.se}     
     
\vspace*{5mm}{\bf Abstract:} 
                                 
\end{center}

\noindent We study (3+1)-dimensional $\cN=4$ supersymmetric Yang-Mills theory on a spatial three-torus. The low energy spectrum consists of a number of continua of states of arbitrarily low energies. Although the theory has no mass-gap, it appears that the dimensions and discrete abelian magnetic and electric 't Hooft fluxes of the continua are 
computable in a semi-classical approximation. The wave-functions of the low-energy states are supported on submanifolds of the moduli space of flat connections, at which various subgroups of the gauge group are left unbroken. 
The field theory degrees of freedom transverse to such a submanifold are approximated by supersymmetric matrix quantum mechanics with 16 supercharges, based on the semi-simple part of this unbroken group. Conjectures about the number of normalizable bound states at threshold in the latter theory play a crucial role in our analysis. In this way, we compute the low-energy spectra in the cases where the simply connected cover of the gauge group is given by 
$\SU (n)$, $\Spin (2 n {+} 1)$ or $\Sp (2 n)$. We then show that the constraints of $S$-duality are obeyed for unique values of the number of bound states in the matrix quantum mechanics. In the cases based on $\Spin (2 n {+} 1)$ and $\Sp (2 n)$, the proof involves surprisingly subtle combinatorial identities, which hint at a rich underlying structure.

\setcounter{equation}{0}
\section{Introduction}
An $\cN = 4$ supersymmetric $(3 + 1)$-dimensional Yang-Mills theory is completely characterized by a gauge group $G$ and a value of the complex parameter 
\beq \label{tau}
\tau = \frac{\theta}{2 \pi} + \frac{i}{g^2} \,,
\eeq
where $\theta$ is the theta angle and $g$ is the coupling constant. The $S$-duality conjecture \cite{Montonen:1977,Osborn:1979} states that this characterization is somewhat redundant: The transformations 
\bea
S : (G, \tau) & \mapsto & ( G^{\vee}, - 1 / \tau) \cr
T : (G, \tau) & \mapsto & (G, \tau + 1) \,,
\eea
both give theories equivalent to the original one. (These expressions are true for simply laced groups. For non-simply-laced cases the situation is more complicated; see e.g.~\cite{Argyres:2006} for a recent discussion.) Here $G^{\vee}$ denotes the GNO or Langlands dual group of $G$ \cite{Goddard:1976}. Some examples of such dualities, that will be studied further in this paper, are:
\beq
\begin{array}{lll}
\underline{G} & \underline{G^{\vee}} & \underline{C} \vspace*{2mm} \cr
\SU(n) &  \SU(n)/\ZZ_n & \ZZ_n \cr
\Spin(2n{+}1) & \Sp(2n)/\ZZ_2 & \ZZ_2 \cr
\Sp(2n) & \Spin(2n{+}1)/\ZZ_2  & \ZZ_2 \,.
\end{array}
\eeq
Here $C$ denotes the center of the simply connected group $G$ (which is isomorphic to the center of the universal covering group of $G^{\vee}$). Whereas the transformation $T$ is manifest in the usual formalism of Yang-Mills theory, the transformation $S$ is still rather mysterious (although by now very well established). The object of the present paper is to give further evidence for it, which we hope could be useful for elucidating its deeper meaning.

We will compare the theories with gauge groups $G$ and $G^\vee$ on a space-time of the form $\mathbb R \times T^3$, where the first factor represents time and the second factor is a three-torus
with a flat metric. The states of the theory are then characterized by two discrete quantum numbers 
\bea
m \in M & \simeq & H^2 (T^3, C) \simeq C^3 \cr
e \in E & \simeq & H^1 (T^3, C) \simeq C^3 .
\eea
The discrete abelian magnetic 't Hooft flux $m$ characterizes the topological class of a gauge bundle over $T^3$. The discrete abelian electric 't Hooft flux $e$ characterizes together with the vacuum angle $\theta$ the transformation properties of the state under "large" gauge transformations with a non-trivial winding in the gauge group \cite{'tHooft:1979}; the former is related to winding around a closed curve in the base manifold, whereas the latter is related to winding over a three-sphere. In a theory with a simply connected gauge group $G$, all states have $m = 0$ but $e$ may take arbitrary values. The gauge group $G^\vee$ of the dual theory is then "of adjoint form", and all states have $e = 0$ while $m$ may take arbitrary values. Intermediate cases, where the gauge group is neither simply connected nor of the adjoint form, give other restrictions on $m$ and $e$. We will be slightly more general, and consider all combinations of $m$ and $e$, although some of them seemingly cannot appear in a gauge theory. $S$-duality then acts according to
\bea
S : (m, e) & \mapsto & (e, -m) \cr
T : (m, e) & \mapsto & (m, e + \Delta) ,
\eea
where the ``spectral flow" $\Delta$ depends on $m$ but not on $e$. These matters are explained in more detail in section two.

In general, the predictions of $S$-duality are difficult to verify, since they relate a weakly coupled theory, in which many quantities are  computable in a semi-classical approximation, 
to a strongly coupled theory, where most quantities are beyond reach. An exception are quantities that are invariant under continuous deformations of the coupling constant, and therefore may be followed in an interpolation between the two regimes. The prototype of such a quantity is the Witten index in a supersymmetric theory with a mass gap, i.e.~the number of bosonic minus the number of fermionic states of zero energy \cite{Witten:1982}. The case of $\cN = 4$ supersymmetric Yang-Mills theory on $T^3$ does not fulfill the standard requirements for this theorem: Supersymmetry ensures that the energy spectrum is non-negative, but there is no mass gap, so the low energy spectrum will consist of continua of non-normalizable states with arbitrarily low energies. Such a continuum is characterized by its dimension $d$ (i.e. the number of continuous parameters needed to label the states) and the discrete quantum numbers $m$ and $e$ introduced in the previous paragraph. (Continua of dimension zero do correspond to normalizable zero energy states.) A priori, it is not clear that this low energy spectrum is invariant under continuous deformations of $\tau$ and the parameters describing the flat metric on $T^3$. However, our results lend strong support to the conjecture that this is indeed the case. It would be interesting to try to find a more rigorous proof of this invariance.

Assuming that the low-energy spectrum is invariant under continuous deformations of the theory, it may be computed semi-classically at weak coupling. This is explained for an arbitrary gauge group $G$ in section three. The main point is that the wave-function of a low energy state is localized at connections with vanishing curvature on a principal $G$ bundle over the spatial $T^3$. The structure of the moduli spaces of such flat connections is known for all simple groups \cite{Witten:1998,Borel:1999,Kac:1999a}. We take this analysis one step further, by studying the submanifolds of these moduli spaces at which various subgroups $H$ of $G$ are left unbroken. 
The abelian factors of $H$ determine the dimensions of the continua of states with wave-functions localized on these submanifolds. The semi-simple factors of $H$ determine the number of continua as follows: The field theory degrees of freedom transverse to such a submanifold may be modelled by matrix quantum mechanics with sixteen supercharges based on these semi-simple factors. (This is the theory obtained by dimensional reduction of $\cN = 4$ supersymmetric Yang-Mills theory in $3 + 1$ dimensions to $0 + 1$ dimensions.) This version of quantum mechanics is believed to have a number (depending on the group) of normalizable zero-energy bound states at threshold. (It should be noted, however, that these states have not yet been rigorously constructed.)  In this way, one may determine the low energy spectrum for any gauge group $G$. As explained above, it consists of a set of continua of states, characterized by their dimensions $d$ and the 't Hooft fluxes $m$ and $e$. The $S$-duality of this spectrum is by no means obvious, though.

In the last two sections, we consider two classes of specific examples, for which we compute the low-energy spectrum and verify that it satisfies the constraints of  $S$-duality. Section four is concerned with the $G = \SU (n)$ case. The Lie algebra of a possible invariant subgroup $H$ is then given by a sum of abelian terms and $\su (k)$ terms. There are strong reasons to believe that $\su (k)$ matrix quantum mechanics has precisely one normalizable state. (This was first predicted on the basis of the duality between type IIA string theory and $M$-theory \cite{Witten:1995}.) It is then not difficult to compute the low-energy spectrum and verify its $S$-duality. (One can also run this argument in reverse: assuming S-duality uniquely fixes the number of bound states in $\su (k)$ matrix quantum mechanics to be precisely one.)

In section five, we consider the cases $G = \Spin (2 n + 1)$ and $G = \Sp (2 n)$. The Lie algebras of the possible unbroken subgroups $H$ may then contain $\so (k)$ and $\spl (2 k)$ terms. Here, there are no well-established predictions for the number of normalizable states in the matrix quantum mechanics. However, a mass deformed version of $\cN = 4$ Yang-Mills field theory, 
known as the $\cN = 1^*$ theory, has a mass gap and a number of vacuum states which is computable. Assuming that the number of bound states in the matrix quantum mechanics can be extracted by taking the massless limit of the $\cN = 1^*$ theories, one is then lead to the conjecture that the number of normalizable bound states at threshold in $\so (k)$ or $\spl (2 k)$ matrix quantum mechanics is related to certain integer partitions~\cite{Kac:1999b}. Our general methods allow for a determination of the low energy spectrum of the $G=\Spin (2 n {+} 1)$ and $G=\Sp (2 n)$  $\cN = 4$ theories on $T^3$. It turns out to be easier to describe the results for all values of $n$ simultaneously, rather than studying a specific group. Provided that the ~conjecture given in \cite{Kac:1999b} for the number of normalizable bound states at threshold in matrix quantum mechanics is true, the predictions of $S$-duality are fulfilled in a surprisingly subtle and intriguing way, with combinatorial identities like Jacobi's  {\it aequatio identica satis abstrusa} making an unexpected appearance. Again, the argument can be reversed, showing that the conjectures in \cite{Kac:1999b} for the number of bound states are the unique choices consistent with $S$-duality.

Our results can be interpreted in different ways. One viewpoint is that they shed light on the intriguing relationships between three unproven (but at this time rather uncontroversial hypotheses): The presumed coupling constant independence of the low-energy spectrum, the question of normalizable states in matrix quantum mechanics, and $S$-duality. Concerning possible generalizations of these results, one would of course like to find a unified description valid for all gauge groups $G$. An obvious first step, which is currently under investigation and on which we hope to report on in the near future, concerns the remaining cases with a simply laced gauge group, i.e. $G = \Spin (2 n)$ and $G = E_{6, 7, 8}$. Hopefully, this can be helpful for understanding the structures underlying the $\cN = 4$ theories, e.g.~a  formulation in terms of a $(5 + 1)$-dimensional $(2, 0)$ theory considered on $T^5 \simeq T^2 \times T^3$, where the geometry of the first factor is related to the parameter $\tau$ (\ref{tau}). The results reported in this paper give us good hope that it should be possible to make further progress along these lines.

We should perhaps also point out that $S$-duality of
the $\cN=4$ super-Yang-Mills theory on $\mathbb{R}{\times}T^3$ has been studied
before in the literature in the context of (M)atrix theory (see
e.g.~\cite{Hacquebord:1997}). However, these studies focus on different
aspects: BPS states, rather than low-energy states are studied, and the
gauge group is $\U(N)$, rather than $\SU(N)$.

\setcounter{equation}{0}
\section{'t Hooft fluxes in non-abelian Yang-Mills theory}
In this section, we will review some algebraic topology aspects of principal $G_{\rm adj}$ bundles over a three-dimensional compact base space $B$. Here $G_{\rm adj} = G / C$ is the adjoint form of a simply connected compact Lie group $G$ with center subgroup $C$. Readers who are less interested in the formal aspects may skip this section without much harm; the most relevant results were summarized in the introduction. A useful reference for this section is \cite{Witten:2000}.

\subsection{The discrete abelian magnetic flux}
For $G$ a simply connected compact Lie group with center $C$, the first few homotopy groups of the quotient group $G_{\rm adj} = G / C$ are given by
\beq
\pi_i (G_{\rm adj}) \simeq \left\{ \begin{array}{ll} 0,  & \; i = 0 \cr C, & \; i = 1 \cr 0, & \; i = 2 \cr \mathbb Z, & \; i = 3 \,. \end{array} \right.
\eeq
It follows that an isomorphism class of a principal $G_{\rm adj}$ bundle $P$ over a compact base space $B$ of dimension less than or equal to four is completely determined by the Stiefel-Whitney class (discrete abelian magnetic 't Hooft flux)
\beq
m = w_2 (P) \in M = H^2 (B, \pi_1 (G_{\rm adj}))  \, ,
\eeq
and the Chern class (instanton number)
\beq
k = c_2 (P) \in H^4 (B, \mathbb Q) \,.
\eeq
(Of course, $m$ or $k$ are trivial if the dimension of  $B$ is less than two or four respectively.) In higher dimensions, there are further invariants, but they will not be needed in the present paper. The classes $m$ and $k$ are not independent: For example, if the center is a cyclic group $C \simeq \mathbb Z_n$, we have that 
\beq
k - \frac{1}{2} (1 - \frac{1}{n}) \bar{m} \cup \bar{m} \in H^4 (B, \mathbb Z) \label{nonintegrality}
\eeq
where $\bar{m} \in H^2 (B, \mathbb Z)$ is a lifting of $m$. (This actually covers all cases, except $G \simeq \Spin (4 k)$ for which $C \simeq \mathbb Z _2 \times \mathbb Z_2$.)

\subsection{The discrete abelian electric flux}
Let $P$ be a principal $G_{\rm adj} = G / C$ bundle over a {\it three}-dimensional compact base space $B$. Let ${\cal G}$ denote the group of bundle automorphisms of $P$ (gauge transformations), and let ${\cal G}_0$ denote the connected component of the identity. It follows from a canonical analysis that physical states must be invariant under ${\cal G}_0$, since the generator of infinitesimal such transformations is weakly zero. But a physical state may transform non-trivially under the discrete abelian quotient group
\beq
\tilde{\Omega} = {\cal G} / {\cal G}_0 \, ,
\eeq
of ``large" bundle automorphisms. The transformation properties under this group is given by a character
\beq
\tilde{e} \in \tilde{E} = {\rm Hom} (\tilde{\Omega}, U (1)) .
\eeq

To describe the structure of the groups $\tilde{\Omega}$ and $\tilde{E}$, we let 
\beq
\Omega = {\rm Hom} (\pi_1 (B), \pi_1 (G_{\rm adj})) 
\eeq
and define the map $r : \tilde{\Omega} \rightarrow \Omega$ by restricting a bundle automorphism of $P$ to closed curves in $B$. (Note that the restriction of $P$ over such a curve is a trivial bundle.) We wish to determine the kernel $\Omega_0 \subset \tilde{\Omega}$ of the map $r$. To this end, we note that a bundle automorphism of $P$ whose restriction to closed curves in $B$ is continuously connected to the identity, may be continuously deformed to a bundle automorphism with support in a small open three-disc $D$ in $B$. Since the restriction of $P$ to $D$ is trivial, such an automorphism is given by a map from a three-sphere $S^3$ (the closure of the disc $D$ with all boundary points identified) to the gauge group $G_{\rm adj}$. The group of homotopy classes of such maps can thus be identified with the kernel of  $r$, i.e.
\beq
\Omega_0 \simeq \pi_3 (G_{\rm adj})  \simeq \mathbb Z.
\eeq

The group $\tilde{\Omega}$ is thus an extension of $\Omega$ by $\Omega_0$:
\beq
0 \rightarrow \Omega_0 \stackrel{i}{\rightarrow} \tilde{\Omega} \stackrel{r}{\rightarrow} \Omega \rightarrow 0 .
\eeq
We wish to describe this extension more precisely. An arbitrary element $\tilde{\omega} \in \tilde{\Omega}$ is mapped by $r$ to an element $\omega \in \Omega$ of some finite order $s$. Exactness of the above sequence then implies that
\beq
(\tilde{\omega})^s = (\Upsilon)^k \label{ks} ,
\eeq
where 
$\Upsilon$ 
is the generator of $\Omega_0$ and $k$ is an integer. The integer $k$ may be determined modulo $s$ as follows: We construct two principal $G_{\rm adj}$ bundles $P_{\tilde{\omega}}$ and $P_\Upsilon$ 
over an auxiliary four-dimensional space $Y = S^1 \times B$ by first extending $P$ over $I \times B$ and then identifying the ends of the interval $I$ to obtain $S^1$ with a twist by $\tilde{\omega}$ and 
$\Upsilon$ respectively. The Chern classes of theses bundles are related as
\beq
s c_2 (P_{\tilde{\omega}}) = k c_2 (P_\Upsilon) .
\eeq
But 
$c_2 (P_\Upsilon) = 1$ 
(where we have identified $H^4 (S^1 {\times} B, \mathbb Q) \simeq \mathbb Q$), and $c_2 (P_{\tilde{\omega}})$ is determined modulo $1$ by the Stiefel-Whitney class $w_2 (P_{\tilde{\omega}})$. (See (\ref{nonintegrality}) for the case when $C \simeq \mathbb Z_n$.) Finally, the latter class is determined by its restriction to $B$, which is given by $w_2 (P)$, and its restrictions to $S^1 \times c$ for an arbitrary loop $c$ in $B$, which are determined by $\omega \in \Omega$. These considerations thus determine the integer $k$ modulo the order $s$ of $\omega \in \Omega$. Note that $k$ modulo $s$ only depends on the image $\omega$ of $\tilde{\omega}$.

Finally, we will describe the relationship between characters of the groups $\Omega_0$, $\tilde{\Omega}$, and $\Omega$. We begin by defining the ``spectral flow'' character
\beq
\Delta  \in E = {\rm Hom} (\Omega, U (1))
\eeq
by evaluating it for an arbitrary element $\omega \in \Omega$:
\beq
\Delta (\omega) = \exp (-2 \pi i k / s) ,
\eeq
where $s$ is the order of $\omega$ and the integer $k$ is determined modulo $s$ as described in the previous paragraph. A character $\tilde{e}$ of the group $\tilde{\Omega}$ now determines a character $e_0$ of the group $\Omega_0$
\beq
e_0 \in  {\rm Hom} (\Omega_0, U (1)) \simeq U (1)
\eeq
and a character $e$ of the group $\Omega$ defined modulo the spectral flow  $\Delta$, i.e.~an equivalence class
\beq
[e] \in E \mod \Delta .
\eeq
The character $e_0$ is given by the restriction of $\tilde{e}$ from $\tilde{\Omega}$ to the subgroup $\Omega_0$. It is determined by the "vacuum angle" $\theta$ defined modulo $2 \pi$ according to 
\beq
e_0 (\Upsilon) = \tilde{e} (\Upsilon) = \exp (i \theta) ,
\eeq
where $\Upsilon$ is the generator of $\Omega_0$. The character $e$ is defined by evaluating it for an arbitrary element $\omega \in \Omega$:
\beq
e (\omega) = \tilde{e} (\tilde{\omega}) \exp (-i \theta k / s) .
\eeq
Here $s$ is the order of $\omega$, $\tilde{\omega}$ is a lifting of $\omega$ to $\tilde{\Omega}$, and the integer $k$ is defined by (\ref{ks}). This is independent of the choice of $\tilde{\omega}$. However, since $\theta$ is only defined modulo $2 \pi$, only the class $[e]$ of $e$ modulo the spectral flow $\Delta$ is really well defined.

\subsection{The action of $S$-duality} \label{sdual}
Physical states are thus characterized by the vacuum angle $\theta$ and the 't Hooft fluxes $m \in M$ and $[e] \in E \mod \Delta$. The groups $E$ and $M$ are (canonically) isomorphic:
\beq
M \simeq E \simeq (C)^{b_1 (B)} ,
\eeq
where $b_1 (B)$ is the first Betti number of $B$. If we choose a representative $e \in E$ of $[e]$, we have seen that the $T$-transformation $\theta \rightarrow \theta + 2 \pi$ acts as 
\beq
(m, e) \rightarrow (m, e + \Delta) \,.
\eeq
The $S$ transformation amounts to the interchange of $E$ and $M$, in the sense that 
\beq
(e, m) \rightarrow (m, -e) \,.
\eeq

In this paper, the base manifold is $B = T^3$ so that $b_1 (B) = 3$. With respect to a basis $(c_1, c_2, c_3)$ of one-cycles of $B$, the 't Hooft fluxes then amount to two triples of elements of $C$:
\bea
m & = & (m_{23}, m_{31}, m_{12})  \in C^3 \cr
e & = & (e_1, e_2, e_3) \in C^3 .
\eea
In an additive notation, these triples transform linearly under the $\SL (3, \mathbb Z)$ mapping class group of $T^3$. If $C \simeq \mathbb Z_n$, this means that $m$ is related by an $\SL (3, \mathbb Z)$ transformation to $(0, 0, u)$, 
where $u =  \mathrm{gcd} (m_{23}, m_{31}, m_{12}, n)$ is the greatest common divisor of $m_{23}$, $m_{31}$, $m_{12}$, and $n$. (Alternatively, one could put $e$ in this form, but in general not both $m$ and $e$ simultaneously.)

\setcounter{equation}{0}
\section{The low energy effective theory}
In this section, we will describe how to compute the low energy spectrum for an arbitrary gauge group of the form $G_{\rm adj} = G  / C$, where $G$ is  a simply connected compact Lie group with center $C$.

At weak coupling, i.e.~$g \ll 1$, the $\cN=4$ Yang-Mills theory is a perturbative quantum field theory with the following fundamental fields: A bosonic connection $A$ on a principal $G_{\rm adj}$-bundle $P$ over space-time,  a bosonic section $\Phi$ of the associated bundle $ad (P)$, and fermionic sections $\Psi_\pm$ of the associated bundles $ad (P) \otimes S_\pm$, where $S_\pm$ are the positive and negative chirality spinor bundles over space-time. In addition to its gauge and space-time symmetries, the theory is invariant under a global $\SU (4) \simeq \Spin (6)$ $R$-symmetry, which commutes with $S$-duality. The fields $A$, $\Phi$, $\Psi_+$, and $\Psi_-$ transform as ${\bf 1}$, ${\bf 6}$, ${\bf 4}$, and ${\bf \bar{4}}$ respectively under $R$-symmetry. We will work in a Hamiltonian formalism in temporal gauge, i.e.~with the time-component of the gauge field put to zero. The above fields can then be viewed as sections of bundles over the spatial three-manifold $B = T^3$. 

By supersymmetry, the spectrum of the Hamiltonian is bounded from below by zero. In general, the precise spectrum depends on the continuous parameters of the theory, e.g. the coupling constant $g$, the theta angle, and the parameters describing the flat metric on $T^3$. But one could hope that the spectrum at arbitrarily low energies should be invariant under smooth deformations of the theory. This is well established for theories with a discrete spectrum below a finite energy gap, but it seems to be true also in the case at hand, where we have continua of states of various dimensions extending down to zero energy. (We consider normalizable states of precisely zero energy as zero-dimensional continua.) 

Our aim is to compute the quantum numbers of this low-energy spectrum, in particular the 't Hooft fluxes $m \in M \simeq C^3$ and $e \in E \simeq C^3$. The Hamiltonian of the theory is a sum of manifestly non-negative terms, all of which must thus be arbitrarily small for the states under consideration. We will consider each of these terms separately. 

\subsection{The magnetic energy}
We consider a principal $G_{\rm adj}$ bundle $P$ over $T^3$ with connection $A$. The magnetic contribution to the energy is proportional to ${\rm tr} (B_i B_i)$, where $B_i = \epsilon_{i j k} F_{j k}$ is given by the spatial components of the curvature $F = d A + A \wedge A = F_{j k} d x^j \wedge d x^k$. A flat connection, for which  $F = 0$, is completely described by the holonomies
\beq
U_i^\prime = P \exp \int_{c_i} A  \,,
\eeq
where $c_i$, $i = 1, 2, 3$ are curves that start and end at a common base-point and whose homotopy classes generate $\pi_1 (T^3) \simeq \mathbb Z^3$. The holonomies $U_i^\prime$ are commuting elements of $G_{\rm adj}$. They may be lifted to elements $U_i$ of $G$ obeying the commutation relations
\beq
U_i U_j = m_{ij} U_j U_i \,, \label{comrel}
\eeq
where we have identified the components $m_{ij} = m_{ji}^{-1}$, of the discrete abelian magnetic flux $m = (m_{23}, m_{31}, m_{12})$ of the bundle $P$ with elements of $C$.  

A gauge transformation (bundle automorphism) acts on the holonomies by conjugation: 
\beq
U_i \rightarrow g U_i g^{-1} ,
\eeq
where $g \in G$ is the parameter of the transformation restricted to the basepoint of the curves $c_i$.
For a given $m \in M$, there is a moduli space ${\cal M} (m)$ of gauge inequivalent triplets $U = (U_1, U_2, U_3)$ of holonomies obeying (\ref{comrel}). The wave-function of a low-energy state with a specific value of $m \in M$ is thus supported on ${\cal M} (m)$.

In general, the moduli space ${\cal M} (m)$ is disconnected:
\beq
{\cal M} (m) = \bigcup_\alpha {\cal M}_\alpha (m)
\eeq
where $\alpha$ labels the different connected components. The rank of the commutant $H \subset G$ of $U$ is locally constant on ${\cal M}$. Denoting its value on the ${\cal M}_\alpha$ component as $r_\alpha$, we have the following formula \cite{Witten:1998,Kac:1999a,Keurentjes:1999a,Borel:1999}:
\beq
\sum_\alpha (r_\alpha + 1) = h^\vee_G ,
\eeq
where $h^\vee_G$ denotes the dual Coxeter number of $G$.

\subsection{The abelian contribution}
At a generic point on the moduli space component ${\cal M}_\alpha (m)$, the Lie algebra $h$ of the commutant $H \subset G$ of $U = (U_1, U_2, U_3)$ is isomorphic to $\ul (1)^{r_\alpha}$, but on subspaces of ${\cal M}_\alpha (m)$, $h$ may be non-abelian. For a given Lie algebra $h$, we let
\beq
{\cal M}^h (m) = \bigcup_a {\cal M}^h_a (m)
\eeq
denote the corresponding subspace of ${\cal M} (m)$, with the spaces ${\cal M}^h_a (m)$ being its connected components. In general, $h$ will contain a semi-simple term $s$ and an abelian term $\ul (1)^r$ for some non-negative integer $r$. We will now analyze the contributions to the energy from the degrees of freedom associated with the abelian $\ul (1)^r$ term. We let ${\rm tr}$ denote the restriction of the Killing form ${\rm Tr}$ of ${\rm Lie \,} G$ to this term. 

The electric contribution to the energy is proportional to ${\rm tr} (E_i E_i)$, where the electric field strength components $E_1$, $E_2$ and $E_3$ are the canonical conjugates of (the $\ul (1)^r$ part of) the holonomies $U_1$, $U_2$, and $U_3$. The wave function of a low-energy state supported on ${\cal M}^h (m)$ must thus be constant on each connected component ${\cal M}^h_a (m)$. (It should be noted that these components are compact.)

The 6 scalar fields $\Phi$ give a contribution to the energy proportional to ${\rm tr} (\Pi \, \Pi)$, where $\Pi$ are the 6 canonical conjugates of the covariantly constant modes of the $\Phi$. (The non-constant modes are not relevant at low energies.) In the quantum theory, there is a continuum of "eigenstates" labelled by the $d = 6 r$ arbitrary real eigenvalues of (the $u (1)^r$ part of) the $\Pi$ operators. (Actually, only the $r = 0$ case would correspond to true normalizable eigenstates.) These states are eigenstates of the ${\rm tr} (\Pi \, \Pi)$ term in the Hamiltonian, which can thus be made arbitrarily small by taking a wave function supported sufficiently close to zero in $\Pi$-space. We refer to this as a rank $r$ continuum of states.

Finally, we need to quantize the covariantly constant modes of the spinor fields $\Psi_\pm$. (Again, the non-constant modes are not relevant at low energies.) These modes are their own canonical conjugates, and give no contribution to the energy. But they give rise to a further finite degeneracy of the low-energy states.

\subsection{The semi-simple contribution}
It remains to consider the degrees of freedom associated with the semi-simple term $s$ in the unbroken Lie algebra $h$. (The degrees of freedom associated with broken generators are massive, and thus  irrelevant at low energies.) It is then convenient to use the canonically normalized variables $X = g^{-1} A$, where $g$ is the coupling constant, instead of the connection $A$. In the weak coupling limit $g \rightarrow 0$, the variables $X$ may then be regarded as non-compact scalars (like the true scalar fields $\Phi$), and the low energy effective theory for the degrees of freedom associated with the semi simple terms $s$ in $h$ is given by $s$ matrix quantum mechanics with 16 supercharges. (The latter theory is most easily described as the dimensional reduction to $0 + 1$ dimensions of the four dimensional $\cN=4$ Yang-Mills theory with the Lie algebra of the gauge group given by $s$.) This matrix quantum mechanics does not have a mass gap, but is believed to have a finite-dimensional linear space $V_s$ of normalizable zero-energy bound states at threshold. 
In terms of the connection $A = g X$, the wave functions of these states are localized on the subspace ${\cal M}^h (m)$ in the $g \rightarrow 0$ limit. In the matrix quantum mechanics, there is also a continuum of states with arbitrarily low energies, but this matches on to the spectrum of states associated with a smaller unbroken semi simple Lie algebra $s^\prime$.

The dimension of the space $V_s$ of bound states in the matrix quantum mechanics is crucial for our discussion, and we will now briefly review the current knowledge concerning this issue: This problem is notoriously difficult since the theory does not have a mass gap and the bound states are at threshold. It is enough to consider the case of a simple Lie algebra $s$; for a semi simple $s$, $V_s$ is given by the tensor product of the spaces corresponding to its simple terms. In the case of  $s \simeq \su (n)$, the duality between IIA string theory and M-theory implies that there should be precisely one such state~\cite{Witten:1995}. Considerations of the Witten index indicate that this is indeed true~\cite{Yi:1997}. Another approach, leading to the same result, is obtained by mass deforming the $\cN = 4$ theory, determining the number of vacua in the resulting $\cN=1^*$ theory, and then taking the massless limit~\cite{Porrati:1997}. For other simple Lie algebras $s$, the situation is less clear. First of all, one does not have a clearcut prediction from string theory. Also, the direct Witten index approach seems much more difficult than for $s \simeq \su (n)$ (see e.g.~\cite{Fischbacher:2003} and references therein). On the other hand, the mass-deformation method can be generalised in a fairly straightforward manner,  and leads to a mathematically completely well-defined problem. The result of this calculation is a linear space $V^*_s$ of vacuum states of dimension \cite{Kac:1999b}
\beq \label{kac}
{\rm dim \,} V^*_s = \left\{ 
\begin{array}{ll}
1 & \mbox{for } s \simeq \su (n) \cr
\mbox{number of partitions of $n$ into distinct odd parts} & \mbox{for } s \simeq \so (n) \cr
\mbox{number of partitions of $2n$ into distinct even parts} & \mbox{for } s \simeq \spl(2n) \\
3,6,11 & \mbox{for } s \simeq \mathrm{e}_{6,7,8} \\
4 & \mbox{for } s \simeq \mathrm{f}_4 \\
2 & \mbox{for } s \simeq \mathrm{g}_2
\end{array}
\right.
\eeq
A priori, it is not clear that $V_s \simeq V^*_s$, but as we will see, $S$-duality lends very strong support to this conjecture. In this context, it should be noted that none of the above approaches to determine $V_s$ relies on $S$-duality. 

For a given unbroken Lie algebra $h$ with semi-simple terms $s$, there is one copy of the vector space $V_s$ associated with each component ${\cal M}^h_a (m)$ of ${\cal M}^h (m)$. Altogether, the non-abelian degrees of freedom may thus be described at low energies by a state in a finite dimensional vector space
\beq
V_h = \bigoplus_a V_s ,
\eeq
given by a direct sum of spaces isomorphic to $V_s$.
 
\subsection{The discrete abelian electric flux}
Let now 
\beq
\omega = (\omega_1, \omega_2, \omega_3) \in \Omega = {\rm Hom} (\pi_1 (T^3), \pi_1 (G_{\rm adj})) \simeq C^3
\eeq
be the restriction of a large gauge transformation $\tilde{\omega} \in \tilde{\Omega}$ to the curves $c_1$, $c_2$, and $c_3$. It acts on the holonomies according to
\beq
U = (U_1, U_2, U_3) \mapsto \omega U = (\omega_1 U_1, \omega_2 U_2, \omega_3 U_3) 
\eeq
(and trivially on the scalar and spinor fields since these are sections of $ad (P)$). Obviously, the commutants in $G$ of $U$ and $\omega U$ agree, so (with $h$ still denoting the Lie algebra of the commutant) we see that $\omega$ acts by permutation on the different connected components ${\cal M}^h_a (m)$ of ${\cal M}^h (m)$. This action induces a linear action of $\omega$ on the finite dimensional vector space $V_h$ introduced in the previous subsection. These transformations for all $\omega \in \Omega$ may be simultaneously diagonalized, i.e. $V_h$ decomposes as a direct sum of one-dimensional subspaces transforming as different characters $e \in E = {\rm Hom} (\Omega, U (1))$.

A compact description of the low energy spectrum is now as follows: For every choice of $m \in M = H^2 (T^3, C)$, $e \in E = {\rm Hom} (\Omega, U (1))$, and $r = 0, 1, \ldots, {\rm rank \, G}$, there is a multiplicity ${\rm mult}_G^r (m, e)$ of rank $r$ continua with discrete fluxes $m$ and $e$. The $S$-duality conjecture implies that
\bea
{\rm mult}_G^r (m, e) & = & {\rm mult}_G^r (m, e + \Delta) \cr
{\rm mult}_G^r (m, e) & = & {\rm mult}_{\overline{G^\vee}}^r (e, -m) ,
\eea
where $\Delta \in E$ is the spectral flow character and $\overline{G^\vee}$ is the simply connected cover of the dual $G^\vee$ of the simply connected group $G$. (In fact, $\overline{G^\vee} = G$ in all cases, except when $G = \Spin (2 n {+} 1)$, for which $\overline{G^\vee} = \Sp (2 n)$ and vice versa.)

\setcounter{equation}{0}
\section{The $G = \SU (n)$ case}
We will now carry out the programme described above for the case when $G=\SU(n)$. The center $C \subset G$ consists of $n \times n$ matrices of the form $c \id_n$ with $c^n = 1$, and is isomorphic to the cyclic group $\mathbb Z_n$. We will identify the magnetic and electric 't Hooft fluxes, as well as (the restrictions to closed curves of) large gauge transformations, with elements of $(\mathbb Z_n)^3$, i.e.
\bea
m & = & (m_{23}, m_{31}, m_{12}) \in (\mathbb Z_n)^3 \cr
e & = & (e_1, e_2, e_3) \in (\mathbb Z_n)^3 \cr
\omega & = & (\omega_1, \omega_2, \omega_3) \in (\mathbb Z_n)^3\, .
\eea
In this notation, the commutation relations of the holonomies $U_i$, $i = 1, 2, 3$ are written as
\beq
U_i U_j = e^{ 2 \pi i m_{ij} / n} \, U_j U_i \,,
\eeq
where $m_{ij} = - m_{ji}$, and transform as
\beq
U_i \mapsto e^{2 \pi i \omega_i / n} U_i \,.
\eeq

A magnetic 't Hooft flux $m = (m_{23}, m_{31}, m_{12})$ is related by a suitable transformation in the $\SL (3, \mathbb Z)$ mapping class group of $T^3$ to the flux
\beq
m = (0, 0, u)  \mod n ,
\eeq
where $u = \gcd (m_{23}, m_{31}, m_{12}, n)$, so we need only consider $m$ of this form. We index the connected components of the moduli space ${\cal M} (m) = \bigcup_\alpha {\cal M}_\alpha (m)$ of flat connections by \cite{Witten:2000}
\beq
\alpha \in \{0, 1, \ldots, v- 1 \} ,
\eeq
where $v = n / u$. On the component ${\cal M}_\alpha (m)$, the holonomies are given by
\beq
U_i = M_i \otimes V^\alpha_i ,
\eeq
where the $M_i$ are arbitrary commuting $\SU (u)$ matrices, and the $V^\alpha_i$ are some fixed $v \times v$ matrices that fulfill the same commutation relations as the $U_i$. We may for example choose
\bea
V^\alpha_1 & = & {\rm diag} \left( e^{i \pi (- v + 1) / v}, e^{i \pi (-v + 3) / v}, \ldots, e^{i \pi (v - 1) / v}  \right) \non \\
V^\alpha_2 & = & {\left( \ba{cccc} 0 & 1 & \cdots & 0 \\
                          \vdots & 0 &\ddots & \vdots \\
                           0& & 0 & 1 \\
                           1 & 0& \hdots & 0
\ea \right) } \\
V^\alpha_3 & = & {\rm diag} \left(e^{2 \pi i \alpha / n}, e^{2 \pi i \alpha / n}, \ldots, e^{2 \pi i \alpha / n} \right) .  \non
\eea

The matrices $M_i$ break $\SU (u)$ to a maximal subgroup $H$. The Lie algebra 
$h$ of $H$ is of the form
\beq
h  \simeq s \oplus  \ul (1)^{k - 1} \simeq \su (u_1) \oplus \ldots \oplus \su (u_k) \oplus \ul (1)^{k - 1} , \label{h}
\eeq
where $k$ is a positive integer and $u_1 + \ldots + u_k = u$. (We see that $r_\alpha = {\rm rank \,} H = u - 1$ independent of $\alpha$,  so that indeed $\sum_\alpha (r_\alpha + 1) = u v = n =  h^\vee_{\SU (n)}$.)  
To describe the structure of the corresponding subspace ${\cal M}^h (m) \subset {\cal M} (m)$ of the moduli space of flat connections, we let $w = \gcd (u_1, \ldots, u_k)$, $(t_1, \ldots t_k) = (u_1 / w, \ldots, u_k / w)$ and $t = t_1 + \ldots t_k = u / w$. The matrices $M_i$ may then be conjugated to the form
\beq
M_i = N_i \otimes \id_w \,,
\eeq
where the $N_i$ are commuting elements of $\U (t)$ that break $\ul (t)$ to the subalgebra $\su (t_1) \oplus \ldots \oplus \su (t_k) \oplus \ul (1)^{k - 1}$ and obey $(\det N_i)^w = 1$. We identify $\det N_i$ with an element of $\mathbb Z_w$, which thus determines a component of the space of $N_i$ matrices. For $i = 1, 2$, this also determines a component of the space of $U_i$ matrices. For $U_3$, we need to take the choice of $\alpha \in \{0, \ldots, v - 1 \}$ into account, so these components are determined by an element of $\mathbb Z_{w v}$.   Altogether, we find that  the connected components of the space ${\cal M}^h (m) = \bigcup_a {\cal M}^h_a (m)$ are indexed by
\beq
a = (a_1, a_2, a_3) \in \mathbb Z_w \times \mathbb Z_w \times \mathbb Z_{w v} .
\eeq
A large gauge transformation  permutes these components as
\beq
a \mapsto a + \bar{\omega} ,
\eeq 
where $\bar{\omega} = (\bar{\omega}_1, \bar{\omega}_2, \bar{\omega}_3) \in \mathbb Z_w \times \mathbb Z_w \times \mathbb Z_{w v}$ is the image of $\omega = (\omega_1, \omega_2, \omega_3) \in (\mathbb Z_n)^3$.

Since the number of normalizable bound states at threshold in $s \simeq \su (u_1) \oplus \ldots \oplus \su (u_k)$ matrix quantum mechanics is (conjectured to be) one, there is a single state $\Psi_a^h$ associated with each component ${\cal M}^h_a (m)$. (As described in section three, this is actually a rank $r = k - 1$ continuum of states, which is further labelled by $6 r$ continuous parameters from the abelian scalar fields and some discrete quantum numbers from the abelian spinor fields.) The action of $\Omega$ on the components induces an action on the space $V^h$ spanned by the $\Psi_a^h$:
\beq
\Psi_a^h \mapsto \Psi_{a + \bar{\omega}}^h \,.
\eeq
For 
\beq
e = (e_1, e_2, e_3) \in t v \mathbb Z_n \times t v \mathbb Z_n \times t \mathbb Z_n \subset (\mathbb Z_n)^3
\eeq
we define the linear combination
\beq
\hat{\Psi}_e^h = \sum_a e^{-2 \pi i (a_1 e_1 + a_2 e_2 + a_3 e_3) / n} \Psi_a^h \,.
\eeq
It is easy to see that it transforms as the character $e$ under large gauge transformations, i.e.
\beq
\hat{\Psi}_e^h \mapsto e^{2 \pi i (\omega_1 e_1 + \omega_2 e_2 + \omega_3 e_3) / n} \hat{\Psi}_e^h ,
\eeq
and that these states together span $V^h$.

We may now describe the low-energy spectrum in an $S$-duality and $\SL (3, \mathbb Z)$ covariant form as follows: Let $t$ be a divisor of $n$, and let $t_1 + \ldots + t_k = t$ be a partition of $t$ into relatively prime parts, i.e.~$\gcd (t_1, \ldots, t_k) = 1$. To these data is associated a rank $r = k - 1$ continuum of low-energy states for every value of $m$ and $e$ of the form
\bea
m & = & t m^\prime \cr
e & = & t e^\prime ,
\eea
where $m^\prime \in (\ZZ_n)^3$ and $e^\prime \in (\ZZ_n)^3$ are subject to the constraint that
\beq
t (m^\prime \times e^\prime) = t (m_{31}^\prime  e_3^\prime  - m_{12}^\prime  e_2^\prime , m_{12}^\prime  e_1^\prime  - m_{23}^\prime  e_3^\prime , m_{23}^\prime  e_2^\prime  - m_{31}^\prime  e_1^\prime)    =  0 .
\eeq
This $S$-duality covariant equation succinctly summarises the low-energy spectrum of the theory. We stress that the $S$-duality of the spectrum was not a priori obvious.

\setcounter{equation}{0}
\section{The $G = \Spin (2n{+}1)$ and $G = \Sp(2n)$ cases}
We will begin by describing some group theoretic facts for $G = \Spin (2n{+}1)$ and $G = \Sp(2n)$. We will then interpret these in terms of orientifolds, before verifying the predictions of $S$-duality. 

\subsection{$G = \Spin (2n{+}1)$}
The gamma matrices $\gamma_i$, $i = 1, \ldots, 2 n + 1$ obey the Clifford algebra
\beq
\{ \gamma_i, \gamma_j \} = 2 \delta_{i j} \id .
\eeq
The product $\gamma_{[i_1} \ldots \gamma_{i_k]} $ is denoted as $\gamma_{i_1 \ldots i_k}$. The Lie algebra $\so (2 n + 1)$ is then spanned by the elements $\gamma_{i j}$, and $G = \Spin (2n{+}1)$ is the corresponding simply connected group obtained by exponentiation. Note that e.g $(\gamma_{12})^2 = - \id$ so that $\exp (\frac{\pi}{2} \gamma_{12}) = \gamma_{12}$.  The center of $G$ is $C = \{\id,- \id \} \simeq \{ 1, -1 \}$.

For $m = (m_{23}, m_{31}, m_{12}) = (1, 1, 1)$, the moduli space of flat connections is of the form
\beq
{\cal M} (m) = {\cal M}_n (m) \cup {\cal M}_{n - 3} (m) ,
\eeq
where we have indexed the connected components by their rank $r_\alpha$. Note that $\sum_\alpha (r_\alpha + 1) = 2 n - 1 = g^\vee_{\Spin (2 n + 1)}$. On the ${\cal M}_n (m)$ component, the holonomies take the form (up to conjugation)
\bea
U_1 & = & \exp( \half[\theta_1^{23} \ga_{23} + \ldots + \theta_1^{2n,2n+1} \gamma_{2 n, 2 n + 1}] ) \non \\
U_2 & = & \exp( \half[\theta_2^{23} \ga_{23} + \ldots + \theta_2^{2n,2n+1} \gamma_{2 n, 2 n + 1}] ) \\
U_3 & = & \exp( \half[\theta_3^{23} \ga_{23} + \ldots + \theta_3^{2n,2n+1} \gamma_{2 n, 2 n + 1}] ) , \non
\eea
and on ${\cal M}_{n-3} (m)$,
\bea
U_1 & = & \ga_{1234} \exp( \half[\theta_1^{89} \ga_{89} + \ldots + \theta_1^{2n,2n+1} \gamma_{2 n, 2 n + 1}] ) \non \\
U_2 & = & \ga_{1357} \exp( \half[\theta_2^{89} \ga_{89} + \ldots + \theta_2^{2n,2n+1} \gamma_{2 n, 2 n + 1}] ) \\
U_3 & = & \ga_{1256} \exp( \half[\theta_3^{89} \ga_{89} + \ldots + \theta_3^{2n,2n+1} \gamma_{2 n, 2 n + 1}] ) . \non 
\eea

For $m=(1,1,-1)$, we similarly have
\beq
{\cal M} (m) = {\cal M}_{n - 1} (m) \cup {\cal M}_{n - 2} (m) .
\eeq
Again, $\sum_\alpha (r_\alpha + 1) = 2 n - 1$. On ${\cal M}_{n - 1} (m)$,
\bea
U_1 & = & \ga_{12} \exp( \half[\theta_1^{45} \ga_{45} + \ldots + \theta_1^{2n,2n+1} \gamma_{2 n, 2 n + 1}] ) \non \\
U_2 & = & \ga_{13} \exp( \half[\theta_2^{45} \ga_{45} + \ldots + \theta_2^{2n,2n+1} \gamma_{2 n, 2 n + 1}] ) \\
U_3 & = & \exp( \half[\theta_3^{45} \ga_{45} + \ldots + \theta_3^{2n,2n+1} \gamma_{2 n, 2 n + 1}] ) , \non
\eea
and on ${\cal M}_{n - 2} (m)$,
\bea
U_1 & = & \ga_{12} \exp( \half[\theta_1^{67} \ga_{67} + \ldots + \theta_1^{2n,2n+1} \gamma_{2 n, 2 n + 1}] ) \non \\
U_2 & = & \ga_{13} \exp( \half[\theta_2^{67} \ga_{67} + \ldots + \theta_2^{2n,2n+1} \gamma_{2 n, 2 n + 1}] ) \\
U_3 & = & \ga_{1234} \exp( \half[\theta_3^{67} \ga_{67} + \ldots + \theta_3^{2n,2n+1} \gamma_{2 n, 2 n + 1}] ) . \non
\eea
In all cases, if we consider the $\theta_i^{2 k , 2 k + 1}$ to be defined modulo $2 \pi$, the $U_i$ are well-defined as elements of $G / C$. (The $U_i$ would be well-defined as elements of $G$, if we consider the $\theta_i^{2 k , 2 k + 1}$ to be defined modulo $4 \pi$.)

Generically the unbroken Lie algebra is $\ul (1)^{r_\alpha}$. However, when
\beq
 (\theta_1^{2 k_1, 2 k_1 + 1}, \theta_2^{2 k_1, 2 k_1 + 1}, \theta_3^{2 k_1, 2 k_1 + 1}) = \pm \ldots = \pm  (\theta_1^{2 k_l, 2 k_l + 1}, \theta_2^{2 k_l, 2 k_l + 1}, \theta_3^{2 k_l, 2 k_l + 1}) , 
\eeq
a term $\ul (1)^l$ is enhanced to $\ul (l) \simeq \su (l) \oplus \ul (1)$. Furthermore, if the common value $\pm (\theta_1, \theta_2, \theta_3)$ of these $l$ triplets belongs to the set
\beq
\Theta = \{ (0, 0, 0), (0, 0, \pi), (0, \pi, 0), (0, \pi, \pi), (\pi, 0, 0), (\pi , 0, \pi), (\pi, \pi, 0), (\pi, \pi, \pi) \} , \label{Theta}
\eeq
there is a further enhancement of $\ul (l)$ to $\so (2 l)$ or $\so (2 l + 1)$ depending on the component in question:
\beq
\begin{array}{ccccc}
\underline{(\theta_1, \theta_2, \theta_3)} & \underline{{\cal M}_n ((1, 1, 1))}  & \underline{{\cal M}_{n - 3} ((1, 1, 1))} & \underline{{\cal M}_{n - 1} ((1, 1, -1))} & \underline{{\cal M}_{n - 2} ((1, 1, -1))} \cr
(0, 0, 0) & \so (2 l + 1) & \so (2 l) & \so (2 l) & \so (2 l + 1) \cr
(0, 0, \pi) & \so (2 l) & \so (2 l + 1) & \so (2 l) & \so (2 l + 1) \cr
(0, \pi, 0) & \so (2 l) & \so (2 l + 1) & \so (2 l + 1) & \so (2 l) \cr
(0, \pi, \pi) & \so (2 l) & \so (2 l + 1) & \so (2 l) & \so (2 l + 1) \cr
(\pi, 0, 0) & \so (2 l) & \so (2 l + 1) & \so (2 l + 1) & \so (2 l) \cr
(\pi, 0, \pi) & \so (2 l) & \so (2 l + 1) & \so (2 l) & \so (2 l + 1) \cr
(\pi, \pi, 0) & \so (2 l) & \so (2 l + 1) & \so (2 l + 1) & \so (2 l) \cr
(\pi, \pi, \pi) & \so (2 l) & \so (2 l + 1) & \so (2 l) & \so (2 l + 1) \cr
\cr
\end{array}
\eeq
So possible unbroken algebras are of the form
\beq
h \simeq s \oplus \ul (1)^r \simeq \so (p_1) \oplus \ldots \oplus \so (p_8) \oplus \su (n_1) \oplus \ldots \oplus \su (n_r) \oplus \ul (1)^r ,
\eeq
where
\beq
p_1 + \ldots p_8 + 2 n_1 + \ldots + 2 n_r = 2 n + 1 .
\eeq
For $m = (1, 1, 1)$ ($m = (1, 1, -1)$), one or seven (three or five) of the $p_1, \ldots, p_8$ are odd.

\subsection{$G = \Sp(2n)$}
The group $G = \Sp(2n)$ consists of all $2n{\times}2n$ unitary matrices $U$ that satisfy $U^T J U = J$, where $J^2=-\id$ and $J^T=-J$. The center is $C = \{\id_{2 n}, - \id_{2 n} \} \simeq \{ 1, -1 \}$. 
We choose the $2 n \times 2 n$ matrix $J$ as
\beq
J = i \sigma_y \otimes \id_n = \, \stackrel{n \; {\mathrm terms}}{\overbrace{i \sigma_y \oplus \ldots \oplus i \sigma_y}} ,
\eeq
where $\sigma_y$ is a Pauli sigma matrix:
\beq
\sigma_x = \left( \begin{array}{cc} 0 & 1 \cr 1 & 0 \end{array} \right)\, , \quad \sigma_y = \left( \begin{array}{cc} 0 & -i \cr i & 0 \end{array} \right)\, , \quad \sigma_z = \left( \begin{array}{cc} 1 & 0 \cr 0 & -1 \end{array} \right) .
\eeq

For $m = (1, 1, 1)$, the moduli space of flat connections has a single connected component of rank $r = n$, so that $r + 1 = n + 1 = g^\vee_{\Sp (2 n)}$. The holonomies are
\bea
U_1 & = & \exp (i \theta_1^1 \si_z) \oplus \ldots \oplus \exp (i \theta_1^n \si_z) \cr
U_2 & = & \exp (i \theta_2^1 \si_z) \oplus \ldots \oplus \exp (i \theta_2^n \si_z) \cr
U_3 & = & \exp (i \theta_3^1 \si_z) \oplus \ldots \oplus \exp (i \theta_3^n \si_z) \,,
\eea
where the $\theta_i^k$ are defined modulo $2 \pi$. 

Generically, the unbroken Lie algebra is $\ul (1)^n$. However, when
\beq
(\theta_1^{k_1}, \theta_2^{k_1}, \theta_3^{k_1}) = \pm \ldots =  \pm (\theta_1^{k_l}, \theta_2^{k_l}, \theta_3^{k_l}) , 
\eeq
a term $\ul (1)^l$ is enhanced to $\ul (l) \simeq \su (l) \oplus \ul (1)$. And if the common value of these triplets is an element of the set $\Theta$ defined in (\ref{Theta}), there is a further enhancement of $\ul (l)$ to $\spl (2 l)$. So possible unbroken algebras are of the form
\beq
h \simeq s \oplus \ul (1)^r \simeq \spl (2 l_1) \oplus \ldots \oplus \spl (2 l_8) \oplus \su (n_1) \oplus \ldots \oplus \su (n_r) \oplus  \ul (1)^r ,
\eeq
where 
\beq
l_1 + \ldots + l_8 + n_1 + \ldots + n_r = n .
\eeq

For $m=(1, 1, -1)$, the holonomies are of the form
\bea
U_1 & = & i \si_z \otimes u_1 \non \\
U_2 & = & i \si_x \otimes u_2 \\
U_3 & = & \id_2 \otimes u_3 \non ,
\eea
where $u_1$, $u_2$, and $u_3$ are commuting elements of $\SO (n)$, $\SO (n)$, and $\mathrm{O} (n)$ respectively. Thus
\beq
{\cal M} (m) = {\cal M}_+ (m) \cup {\cal M}_- (m) ,
\eeq
where the two components are distinguished by $\det u_3 = \pm 1$. For $n$ even, ${\cal M}_+ (m)$ has rank $r_+ = n / 2$ and holonomies
\bea
U_1 & = & i \si_z \otimes \left( \exp (i \theta_1^1 \si_y) \oplus \ldots \exp (i \theta_1^{n / 2} \si_y) \right) \cr
U_2 & = & i \si_x \otimes \left( \exp (i \theta_2^1 \si_y) \oplus \ldots \exp (i \theta_2^{n / 2} \si_y) \right) \cr
U_3 & = & \id_2 \otimes \left( \exp (i \theta_3^1 \si_y) \oplus \ldots \exp (i \theta_3^{n / 2} \si_y) \right) ,
\eea
whereas ${\cal M}_- (m)$ has rank $r_- = n / 2 - 1$ and
\bea
U_1 & = & i \si_z \otimes \left( \exp (i \theta_1^1 \si_y) \oplus \ldots \exp (i \theta_1^{n / 2 - 1} \si_y) \oplus  \id_2 \right) \cr
U_2 & = & i \si_x \otimes \left( \exp (i \theta_2^1 \si_y) \oplus \ldots \exp (i \theta_2^{n / 2 - 1} \si_y) \oplus \id_2 \right) \cr
U_3 & = & \id_2 \otimes \left( \exp (i \theta_3^1 \si_y) \oplus \ldots \exp (i \theta_3^{n / 2 - 1} \si_y) \oplus \si_z \right) .
\eea
Thus, $(r_+ + 1) +  (r_- + 1) = n + 1 = g^\vee_{\Sp (2 n)}$. For $n$ odd, both ${\cal M}_+ (m)$ and ${\cal M}_- (m)$ have rank  $r_+ = r_- = (n - 1) / 2$, so that again $(r_+ + 1) +  (r_- + 1) = n + 1 = g^\vee_{\Sp (2 n)}$. The holonomies are
\bea
U_1 & = &  i \si_z \otimes \left( \exp (i \theta_1^1) \si_y \oplus \ldots \exp (i \theta_1^{(n - 1) / 2} \si_y ) \oplus \id_1 \right) \cr
U_2 & = &  i \si_x \otimes \left( \exp (i \theta_2^1\si_y ) \oplus \ldots \exp (i \theta_2^{(n - 1) / 2} \si_y ) \oplus \id_1 \right) \cr
U_3 & = &  \id_2 \otimes \left( \exp (i \theta_3^1 \si_y ) \oplus \ldots \exp (i \theta_3^{(n - 1) / 2} \si_y ) \oplus \id_1 \right) ,
\eea
and
\bea
U_1 & = &  i \si_z \otimes \left( \exp (i \theta_1^1 \si_y ) \oplus \ldots \exp (i \theta_1^{(n - 1) / 2} \si_y ) \oplus \id_1 \right) \cr
U_2 & = & i \si_x \otimes \left(  \exp (i \theta_2^1 \si_y ) \oplus \ldots \exp (i \theta_2^{(n - 1) / 2} \si_y ) \oplus \id_1 \right) \cr
U_3 & = &  \id_2 \otimes \left( \exp (i \theta_3^1 \si_y ) \oplus \ldots \exp (i \theta_3^{(n - 1) / 2}  \si_y) \oplus -\id_1 \right) ,
\eea
respectively. In all cases, the $\theta_1^k$ and $\theta_2^k$ are defined modulo $\pi$, whereas $\theta_3^k$ is defined modulo $2 \pi$.

Generically, the unbroken Lie algebra is $\ul (1)^{r_\alpha}$. Again, enhancement of a term $\ul (1)$ to $\ul (l) \simeq \su (l) \oplus \ul (1)$ occurs when
\beq
(2 \theta_1^{k_1}, 2 \theta_2^{k_1}, \theta_3^{k_1}) = \pm \ldots =  \pm (2 \theta_1^{k_l}, 2 \theta_2^{k_l}, \theta_3^{k_l})  ,
\eeq 
and further enhancement occurs if the common value $(2 \theta_1, 2 \theta_2, \theta_3)$ of these triplets is an element of the set $\Theta$: 
\beq
\begin{array}{ccccc}
\underline{(2 \theta_1, 2 \theta_2, \theta_3)} & \underline{{\cal M}_{+, n / 2}}  & \underline{{\cal M}_{-, n / 2 - 1}} & \underline{{\cal M}_{+, (n - 1) / 2}} & \underline{{\cal M}_{-, (n - 1) / 2}} \cr
(0, 0, 0) & \so (2 l) & \so (2 l + 1) & \so (2 l + 1) & \so (2 l) \cr
(0, 0, \pi) & \so (2 l) & \so (2 l + 1) & \so (2 l) & \so (2 l + 1) \cr
(0, \pi, 0) & \spl (2 l) & \spl (2 l) & \spl (2 l) & \spl (2 l) \cr
(0, \pi, \pi) & \spl (2 l) & \spl (2 l) & \spl (2 l) & \spl (2 l) \cr
(\pi, 0, 0) & \spl (2 l) & \spl (2 l) & \spl (2 l) & \spl (2 l) \cr
(\pi, 0, \pi) & \spl (2 l) & \spl (2 l) & \spl (2 l) & \spl (2 l) \cr
(\pi, \pi, 0) & \spl (2 l) & \spl (2 l) & \spl (2 l) & \spl (2 l) \cr
(\pi, \pi, \pi) & \spl (2 l) & \spl (2 l) & \spl (2 l) & \spl (2 l) .
\end{array}
\eeq
So for $n$ even, possible unbroken algebras are of the form
\beq
h \simeq s \oplus u (1)^r \simeq \left\{ 
\begin{array}{l} 
\so (2 k_1) \oplus \so (2 k_2) \oplus \spl (2 l_1) \oplus \ldots \oplus \spl (2 l_6) \cr
\oplus \, \su (n_1) \oplus \ldots \oplus \su (n_r) \oplus \ul (1)^r \cr
\;\;\;\;\;\;\;\;\;\;\;\; \mathrm{or} \cr 
\so (2 k_1 + 1) \oplus \so (2 k_2 + 1) \oplus \spl (2 l_1) \oplus \ldots \oplus \spl (2 l_6) \cr
\oplus \, \su (n_1) \oplus \ldots \oplus \su (n_r) \oplus \ul (1)^r ,
\end{array}
\right.
\eeq
whereas for $n$ odd,
\bea
h \simeq s \oplus \ul (1)^r & \simeq & \so (2 k_1 + 1) \oplus \so (2 k_2) \oplus \spl (2 l_1) \oplus \ldots \oplus \spl (2 l_6) \cr
& & \oplus \, \su (n_1) \oplus \ldots \oplus \su (n_r) \oplus \ul (1)^r .
\eea
In both cases
\beq
k_1 + k_2 + l_1 + \ldots + l_6 + n_1 + \ldots + n_r  = r_\alpha ,
\eeq
where $r_\alpha$ is the rank of the component in question.

\subsection{Correspondence with orientifolds}

Above we gave a general discussion of the holonomies associated with the various moduli spaces. There is convenient way to keep track of the combinatorics of which holonomies are possible. As this alternative method is quite powerful, we will describe it briefly.

It can be shown that the possible holonomies are in one-to-one correspondence with certain D-brane configurations on an orientifold of an auxiliary three torus. The particulars of these orientifolds have been worked out in \cite{Witten:1998,Keurentjes:2000}. 
Some of these orientifolds are inconsistent as string theories (have anomalies), but that is not relevant here as we are only interested in the field theory limit. For a discussion of string theory orientifolds, see \cite{deBoer:2001}.

The orientifold action in question is a $\ZZ_2$ action on the auxiliary three torus and has eight fixed points (see \cite{Witten:1998,Keurentjes:2000} for more details). At each of these eight fixed points sits an orientifold plane ($O$ plane). These can be of two types, $O^-$ and $O^+$. 
Exactly which types occur depend on the gauge group and the values of $m_{ij}$. On these orientifolds one then places a certain number of D-branes, the number of which depend on the particular case and will be described below. As usual, enhanced gauge symmetry is obtained when several branes are on top of each other. 
Away from the orientifold planes one gets enhanced $\ul(n)$ symmetry when $n$ branes are on top of each other (the $n$ mirror branes are also coincident). When $2n$ branes ($n$ branes and their $n$ mirrors) are on top of an $O^+$ plane one gets $\spl(2n)$ enhancement and when $n$ branes are located at an $O^-$ plane one gets $\so(n)$ symmetry. The possible gauge groups one obtains in this way are in one-to-one correspondence with the possible unbroken gauge groups in the gauge theory.   Note that the $O^-$ planes can support single (`fractional') branes which are stuck at the $O^-$ plane. From the above discussion it follows that if we want no abelian $\ul(1)$ factors in the unbroken gauge symmetry then all the branes need to lie at the eight orientifold planes.  Here we will focus on these special points in the moduli space. But one can also discuss the other points in the moduli space using this language.

\paragraph{$\underline{\Spin(2n{+}1)}$}
For $\Spin(n)$ with $m=(1,1,1)$ the relevant $\ZZ_2$ orientifold contains eight  $O^-$ planes located at the eight fixed points of the  $\ZZ_2$ action~\cite{Witten:1998}. One may visualise the eight $O$ planes as lying at the corners of a cube; this picture will be useful later. Each of the  $O^-$ planes can support single (`fractional') D-branes. 
Two fractional branes are equivalent to a brane-mirror pair and can be moved off the $O^-$ plane.
In order for the gauge group to be $\Spin(n)$ (and not just $\SO(n)$ or $\Pin(n)$) one requires that the first and second Stiefel-Whitney classes vanish \cite{Witten:1998}. The solution to this requirement for $\Spin(2n+1)$ shows that there are two components of the moduli space~\cite{Witten:1998}: either one fixed orientifold plane (`the origin') is occupied by a fractional brane, or the other seven orientifold planes are. In addition, in the former case one also has $2n$ branes ($n$ brane-mirror pairs), and in the latter case $2n{-}6$ branes, which should be distributed among the eight orientifolds planes (in pairs).

The brane configurations are translated into expressions for the holonomies as follows. A fractional brane located at one of the eight orientifold planes corresponds to the following eight three vectors (one may visualise these eight possibilities as the corners of a cube):
\be \label{Spinfrac}
\ba{cccccccc}
1 &\quad \ga_k &\quad 1   &\quad \ga_k   &\quad 1 &\quad \ga_k &\quad 1   &\quad \ga_k \\
1 &\quad 1   &\quad \ga_k &\quad \ga_k   &\quad 1 &\quad 1   &\quad \ga_k &\quad \ga_k \\
1 &\quad 1   &\quad 1   &\quad 1 &\quad \ga_k  &\quad \ga_k &\quad \ga_k &\quad \ga_k \\
\ea
\ee
Here $\ga_k$ denotes one of the usual gamma matrices. The index $k$ is correlated with the index labelling the various branes (here we view a brane-mirror pair as two fractional branes). 
The first entry in (\ref{Spinfrac})  corresponds to the special orientifold plane, `the origin', used above to describe the two components of the moduli space. 
The building blocks (\ref{Spinfrac}) can be used to construct the holonomies. The method should be clear from the following example. 
For instance, if (in $\Spin(7)$) there are 3 branes ($k=1,2,3$) at the first orientifold plane and four ($k=4,5,6,7$) at the fifth one gets:
\bea \label{ex1}
U_1 &=&\pm 1 \non \\
U_2 &=& \pm 1       \\
U_3 &=& \ga_4\ga_5\ga_6\ga_7 \,. \non
\eea
Here we have indicated that in addition to the above rules one can also have overall $\pm$ factors in front of the three holonomies. Some of these may be removable by gauge transformations. In the above example the sign ambiguities in front of $U_3$ can be removed by conjugation with the $\Spin(7)$ group element $\ga_1\ga_4$. More generally, if only one $O$ plane is occupied no signs can be removed, if two are occupied (by at least one brane each) one sign can be removed, and if three are occupied, two signs can be removed. If four are occupied and lie on a plane intersecting the cube with the eight corners containing the $O$ planes, two signs can be removed, otherwise all three signs can be removed. 
If five or more points are occupied all signs can be conjugated away.  This discussion about which signs can be removed by gauge transformations is important for the determination of which values of $e=(e_1,e_2,e_3)$ that occur. Acting with a large gauge transformation of the form $\om=(\om_1,\om_2,\om_3)$ where $\om_i$, say, is the non-trivial element of the centre, changes the sign of $U_i$. Therefore, if the sign change can be undone by a gauge transformation one finds that the corresponding wave function has $e_i=1$. In particular, in the second component of the moduli space all signs can be conjugated away and hence all states have $e=(1,1,1)$. In the above example (\ref{ex1}) there are four states and the corresponding values of $e$ are: $(1,1,1)$, $(-1,1,1)$,  $(1,-1,1)$, and $(-1,-1,1)$.
 
It is easy to see that the gauge enhancement that one obtains from $n$ branes at an $O^-$ plane is generated by the Lie algebra elements $\ga_{ij}$, where $i,j$ lie in the corresponding index range. In the example above (\ref{ex1}), $\ga_{ij}$ ($i,j=1,\ldots,3$) generate $\so(3)$, whereas one gets $\so(4)$ from $\ga_{ij}$ ($i,j=4,\ldots,7$).

The next case to consider is $\Spin(2n{+}1)$ with $m=(1,1,-1)$ (or any of its six images under the $\SL(3,\ZZ)$ symmetry). 
Since we can no longer lift to $\Spin(2n{+}1)$ because of the obstruction caused by $m$  we want the second Stiefel-Whitney class to be non-vanishing (note that we will continue to write the holonomies in the covering group $\Spin(2n{+}1)$). This case is therefore often referred to as the case `without spin structure'. We still require the first Stiefel-Whitney class to vanish though since otherwise we get $\mathrm{O}(2n{+}1)$ instead of $\SO(2n{+}1)\equiv \Spin(2n{+}1)/\ZZ_2$. The solution in the orientifold language is that (see e.g.~\cite{Keurentjes:2000}) three orientifold planes need to be occupied by a fractional brane. The three $O$ planes need to be such that, when viewing the positions of the eight $O$ planes as the corners of a cube, neither can be at the origin and they need to lie in a plane through the origin.  
The various possible planes one selects are permuted by $\SL(3,\ZZ)$ and correspond to the different possible values of $m$. The complement of five fractional branes also gives a solution and there are therefore two components of the moduli space.  On the first (second) component of the moduli space one has in addition to the fractional branes also $2n{-}2$ ($2n{-}4$) branes leading to  rank $n{-}1$ ($n{-}2$).
In particular, when $m = (1,1,-1)$, the $O$ planes corresponding to columns two through four (or one and five through eight) in (\ref{Spinfrac}) should have fractional branes; in other words, the total number of branes at these $O$ planes should be odd. 
As an example (in $\Spin(7)/\ZZ_2$) consider three branes at the second $O$ plane, three at the third and one at the fourth giving:
\bea \label{ex2}
U_1 &=& \ga_1 \ga_2 \ga_3   \ga_7 \non \\
U_2 &=& \ga_4  \ga_5 \ga_6 \ga_7  \\
U_3 &=& \pm 1 \non \,.
\eea
Note that $U_1 U_2 = -U_2 U_1$ as expected. This configuration has $\so(3)\oplus\so(3)$ gauge enhancement. We have indicated in (\ref{ex2}) that the sign ambiguities in front of $U_1$ and $U_2$ can be removed by gauge transformations. The associated two states therefore have $e=(1,1,1)$ and $(1,1,-1)$.

\paragraph{$\underline{\Sp(2n)}$} 

In the case of $\Sp(2n)$ with $m=(1,1,1)$ the relevant orientifold contains eight $O^+$ planes \cite{Keurentjes:2000} on which $2n$ D-branes should be distributed. Only brane-mirror pairs can be located at the orientifold planes and each such pair correspond to one of the following eight building blocks: 
\be \label{Sppair}
\ba{rrrrrrrr}
\id_2 &\quad -\id_2  &\quad \id_2   &\quad -\id_2   &\quad \id_2 &\quad -\id_2 &\quad \id_2   &\quad -\id_2 \\
\id_2 &\quad \id_2   &\quad -\id_2 &\quad -\id_2   &\quad  \id_2 &\quad \id_2  &\quad -\id_2 &\quad -\id_2 \\
\id_2 &\quad \id_2   &\quad \id_2   &\quad \id_2 &\quad -\id_2  &\quad -\id_2 &\quad -\id_2 &\quad -\id_2 \\
\ea
\ee
There is only one component of the moduli space and the maximal rank of the unbroken gauge group is $n$. The large gauge transformations act by permuting the above eight possibilities. Depending on the exact configuration, this action may be possible to undo by a gauge transformation. We will return to this point below. As an explicit example, consider the $\Sp(6)$ theory with two branes at the first $O$ plane and four at the sixth leading to:
\bea \label{ex3}
U_1 &=& \mathrm{diag}(\id_2,-\id_4) \non \\
U_2 &=&  \mathrm{diag}(\id_2,\id_4)  \\
U_3 &=& \mathrm{diag}(\id_2,-\id_4)  \non \,.
\eea
 The unbroken gauge symmetry is $\spl(2)\oplus\spl(4)$. Under the action of large gauge transformations, the states corresponding to the above holonomies mix with the states corresponding to the other 55 ways to place two branes at one orientifold plane and four at another. Diagonalising the action leads to the result that, within this class of states, each of the eight possible values of $e$ occurs seven times.

When $m$ is non-trivial, e.g. $m=(1,1,-1)$, the orientifold has two $O^-$planes and six  $O^+$ planes \cite{Keurentjes:2000}. 
The building blocks of the holonomies are for example as follows. Fractional branes located at the two $O^-$ planes correspond to 
\be \label{Spfrac}
\ba{cc}
i\si_z   &\quad i\si_z \\
i\si_x &\quad i\si_x \\
\id_2 &\quad  -\id_2
\ea
\ee
Brane-mirror pairs at the six $O^+$ planes correspond to:
\be \label{Sppair2}
\ba{cccccc}
i\si_z \otimes i \si_y  & i\si_z \otimes i \si_y  & i\si_z \otimes i \si_y & i\si_z \otimes i \si_y & i\si_z \otimes \id_2  &i\si_z \otimes \id_2  \\
i\si_x\otimes i\si_y & i\si_x\otimes i\si_y & i\si_x \otimes \id_2 & i\si_x \otimes \id_2 &i\si_x\otimes i\si_y & i\si_x \otimes  i\si_y \\
\id_4 & -\id_4 &\id_4 & -\id_4 &\id_4 & -\id_4 
\ea
\ee
Note that combining two fractional branes of one of the above types (\ref{Spfrac}) give the remaining two types, completing the pattern in (\ref{Sppair2}). 

There are two components of the moduli space. When $n$ is even, either both $O^-$ planes are each occupied by a fractional brane, or neither of them are. In addition, one has $n{-}2$ ($n$) branes which should be placed (in pairs) at the eight $O$ planes. Note that the total number of branes is only {\it half} the number compared to the $m=(1,1,1)$ case. When $n$ is odd,  one or the other of the $O^-$ planes is occupied by a fractional brane. In addition, one has $n{-}1$ branes which should be placed (in pairs) at the eight $O$ planes.

From the above expressions (\ref{Spfrac}) and (\ref{Sppair2}), it is clear that the signs in front of $U_1$ or $U_2$ can always be removed by conjugation with a suitable group element. 

As an example, consider the $\Sp(6)$ theory with three branes at one of the 
two $O^-$-planes leading to:
\be \ba{rclcl} \label{ex4}
U_1 &=&\mathrm{diag}(i\si_z,i\si_z,i\si_z ) &=& i\si_z \otimes \id_3  \\
U_2 &=& \mathrm{diag}(i\si_x,i\si_x,i\si_x) &=&  i\si_x \otimes \id_3   \\
U_3 &=& \pm \mathrm{diag}(\id_2, \id_2,\id_2) &=& \pm \id_2 \otimes \id_3\,.
\ea \ee
The unbroken gauge symmetry is $\so(3)$, since any Lie algebra element 
of the form $\id_2 \otimes A$ commutes with the above expressions and 
such elements belong to the $\spl(6)$ Lie algebra provided that $A^T = -A$ (in this sector $J=i\si_y\otimes\id_3$).
The $\pm$ ambiguity in (\ref{ex4}) distinguishes the two $O^-$ planes, which consequently are interchanged by large gauge transformations. (When viewing the orientifold planes as the corners of a cube, the action of $\omega_3$ has a geometric meaning: it simply corresponds to reflection in the plane parallel to $z=0$ that divides the cube into two halves.) 
The two states corresponding to (\ref{ex4}) therefore have $e=(1,1,1)$ and $e=(1,1,-1)$. 

\subsection{$S$-duality}
We are now ready to investigate the $S$-duality properties of the low-energy spectrum. As several ingredients enter the analysis, it is helpful to first consider some explicit examples before moving on to the general case. In particular, we will begin by analyzing the zero-dimensional continua, where the unbroken Lie algebra has no abelian terms.

\paragraph{$\underline{\Spin(2n+1)}$}

When $m=(1,1,1)$, there is a trick one can use to evaluate the spectrum of values of $e$: It must decompose into representations of $\SL(3,\ZZ)$. The relevant representations are 
\bea
R_1 & = & \{(1,1,1) \} \cr
R_7 & = & \{(1,1,-1), (1, -1, 1), (1, -1, -1), (-1, 1, 1), \cr
& & (-1, 1, -1), (-1, -1, 1), (-1, -1, -1) \} .
\eea
Furthermore, the number of states with $e=m=(1,1,1)$ should be the same in the $\Spin(2n+1)$ and $\SO(2n+1) = \Spin(2n+1)/\ZZ_2$ theories. The number of states in the $\SO(2n+1)$ theory  is easy to determine: the overall sign ambiguities in the $U_i$'s are absent and all states belong to $R_1$, i.e. they have $e=(1,1,1)$. As an example we list the result for $G = \Spin(7)$:
\be
\begin{tabular}{cc}
 \underline{$h$} & \underline{$e$} \\
 $\so(7)$ & $R_1\oplus R_7$  \\
 $\so(6)$ & $7\, R_1 \oplus 3\, R_7  $  \\
 $\so(4)\oplus\so(3)$ & $7\, R_1 \oplus 3\, R_7  $ \\
 $\emptyset$ & $R_1$.  
\end{tabular}
\ee
 
When $m = (1,1,-1)$ the situation is similar. It is important to note that, in the orientifold setup, only the component with three fractional branes can lead to non-trivial $e$'s (cf.~the discussion in the previous subsection). Furthermore, it is only in the $U_3$ direction that one can have a non-trivial $e_i$. If, on the first component of the moduli space, one only occupies the three $O$ planes containing the fractional branes, one gets as many $e=(1,1,1)$ as $e=(1,1,-1)$ states. With a further $O$ plane occupied, one gets an $e=(1,1,-1)$ state plus an $e=(1,1,1)$ state when the fourth $O$ plane is at the origin, and only an $e=(1,1,1)$ state if one of the other $O$ planes is occupied. Using these results we get in the  $G = \Spin(7)$ case:
\be
\begin{tabular}{cc}
\underline{$h$} & \underline{$e$} \\
 $\so(5)$ & $3\,(1,1,1)\oplus 3\,(1,1,-1)$ \\
 $\so(3) \oplus \so(3)$ & $3\,(1,1,1)\oplus 3\,(1,1,-1)$ \\
 $\so(4)$ & $5\,(1,1,1)\oplus \,(1,1,-1)$  \\
 $\so(3)$ & $5 (1,1,1)$
\end{tabular}
\ee

\paragraph{$\underline{\Sp(2n)}$}  

The determination of the possible unbroken subgroups and values of $e$ is similar to the $\Spin(2n{+}1)$ case. As an example, when $G = \Sp(6)$ and $m=(1,1,1)$, we get:
\be
\begin{tabular}{cc}
\underline{$h$} & \underline{$e$} \\
 $\spl(6)$ & $R_1\oplus R_7$  \\
 $\spl(4)\oplus\spl(2)$ & $7\,R_1 \oplus 7\, R_7$  \\
 $\spl(2)\oplus\spl(2)\oplus\spl(2)$ & $ 7\,R_1 \oplus 7\, R_7$ 
\end{tabular}
\ee
When $m=(1,1,-1)$ only $e_3$ can be non-trivial as we saw before. For $G = \Sp(6)$ we find:
\be
\begin{tabular}{cc}
\underline{$h$} & \underline{$e$} \\
 $\so(3)$ & $(1,1,1)\oplus(1,1,-1)$ \\
 $\spl(2)$ & $6\,(1,1,1)\oplus 6\, (1,1,-1)$
\end{tabular}
\ee

As we described in earlier sections, $S$-duality involves the transformation $\!(e,m) \!\mapsto (m,-e)$. 
It is sufficient to consider states with $m=(1,1,m_3)$ and $e=(1,1,e_3)$ and thus there are four values of $(m_3,e_3)$ to investigate. 
Supersymmetric quantum-mechanical matrix systems associated with groups of rank less than or equal to two have a single bound state. 
(This could be inferred by repeating the above analysis for the $G = \Spin(5), \Sp(4), \Spin(3), \Sp (2)$ cases). 
The above tables then show that $S$-duality determines the number of bound states for supersymmetric $\so(7)$, $\so(6)$, and $\spl(6)$ matrix quantum mechanics to be 1, 1,  and 2 respectively. 

There is one important lesson to be learned from this example: As we increase the rank, the new unknown parameters are the number of bound states for $\spl(2n)$, $\so(2n{+}1)$ and $\so(2n)$ matrix quantum mechanics, but as $S$-duality imposes at least three independent equations relating them, they will all be determined. Thus we have a consistent iterative procedure which can be continued to arbitrarily high rank. In particular, we see that if an $S$-dual solution exists, it is unique. However, it is clear that the above method involving listing all possible unbroken subgroups quickly becomes very cumbersome. To find a more efficient approach, we will simply use the Kac-Smilga conjectures (\ref{kac}) and check if they are consistent with $S$-duality. 

We recall from the previous subsection that for $G=\Spin(2n{+}1)$  the unbroken gauge symmetries without abelian terms are always of the form $\bigoplus_{i=1}^{8} \so(n_i)$, where $\sum_{i=1}^{8} n_i =2n{+}1$. On the different components of the moduli spaces a certain number of the $n_i$'s are odd and the rest even. The Kac-Smilga conjecture states that the number of bound states in the supersymmetric $\so(n)$ matrix quantum mechanics is equal to the number of ways to partition $n$ into distinct odd parts, cf.~(\ref{kac}). 
Combining these results gives  the total number of bound states, but it seems like a rather complicated combinatorial quantity to calculate. However, we  are not really interested in the actual number of bound states, but only want to know if it agrees with the corresponding $G=\Sp(2n)$ quantity. Problems such as this one are common in the theory of partitions, and one can use the powerful language of generating functions to simplify the analysis. We first note that the number of ways to partition $n_i$ into distinct odd parts is given by the coefficient of $q^{n_i}$ in the (formal) power series expansion of 
\beq
P(q) = \prod_{k=1}^{\infty}(1+q^{2k-1}) . 
\eeq
It follows that the number of ways to partition $2n_i$ into distinct odd parts is given by the coefficient of $q^{2 n_i}$ in
\beq
P_e(q)  = \half[P(q)+P(-q)] ,
\eeq
and the number of ways to partition $2n_i{+}1$ into distinct odd parts is given by the coefficient of $q^{2 n_i + 1}$ in
\beq
P_o(q) = \half[P(q)-P(-q)] .
\eeq 
We can now write down the generating functions for the number of bound states in the $G=\Spin(2n{+}1)$ theories for  fixed $(m,e)$. By using the $\SL(3,\ZZ)$ symmetry it is sufficient to restrict to the cases where only $e_3$ and $m_3$ can be non-trivial. The result is:
\bea \label{songen}
\mbox{\underline{$(e_3,m_3)$}} && \mbox{\underline{Generating function}} \non \\
\!(1,1)     && \!\!\! P_o P_e^7 + P_e P_o^7 = {\ts \frac{1}{128} }[P(q)^8 {-} P(-q)^8] + {\ts \frac{7}{64} }[P(q)^6 P(-q)^2 {-} P(-q)^6 P(q)^2] \non \\
\!(1,-1)     && \!\!\!P_o^3 P_e^5 + P_e^3 P_o^5 = {\ts \frac{1}{128} }[P(q)^8 {-} P(-q)^8] - {\ts \frac{1}{64} }[P(q)^6 P(-q)^2 {-} P(-q)^6 P(q)^2] \non \\
\!(-1,1)     && \!\!\!P_o P_e^3 = {\ts \frac{1}{16} }[P(q)^4 {-} P(-q)^4] + {\ts \frac{1}{8} } [P(q)^3 P(-q) {-} P(-q)^3 P(q)]  \\
\!(-1,-1)     && \!\!\!P_o^3 P_e = {\ts \frac{1}{16} }[P(q)^4 {-} P(-q)^4] - {\ts \frac{1}{8} } [P(q)^3 P(-q) {-} P(-q)^3 P(q)] \non \,.
\eea
The fact that only four $P$'s occur when $e_3\neq 1$ follows from the fact that when more than four orientifold planes are occupied one necessarily has $e=(1,1,1)$.  In summary, the coefficient in front of $q^{2n+1}$ in the formal Taylor expansion of the expressions in (\ref{songen}) gives the number of bound states in the $G=\Spin(2n{+}1)$ theories with the corresponding $(e_3,m_3)$ values.

On the $G=\Sp(2n)$ side, a similar analysis can be performed. When $m=(1,1,1)$, possible unbroken Lie algebras without abelian terms are of the form $\bigoplus_{i=1}^{8} \spl(2n_i)$ with $\sum_{i=1}^8 2n_i = 2n$. The Kac-Smilga conjecture states that the number of bound states in the $\spl(2n_i)$ theory is equal to the number of ways one can partition $2n_i$ into distinct even integers, which in turn equals the coefficient of $q^{2n_i}$ in
\beq
Q(q) = \prod_{k=1}^{\infty} (1+q^{2k}) .
\eeq
Large gauge transformations act by permuting the eight $O^+$ planes. If this action were free, the number of states with $e=(1,1,1)$ in the $\Sp(2n)$ theory would be given by the coefficient of $q^{2n}$ in $\frac{1}{8} Q(q)^8$. However, the action is not free, as for certain configurations it can be undone by a gauge transformation. More precisely, if the states arise from partitions occurring in pairs, then the action is not free and there are more states with $e=(1,1,1)$ than one would naively expect. The generating function for such `paired' states is $Q(q^2)^4$. Taking this into account, one is lead to the expressions that are summarised in the table below, where we also multiplied the generating function by an overall factor of $q$ to facilitate the later comparison with the $\Spin(2n{+}1)$ expressions. Note that with $m=(1,1,1)$, summing over all possible $e$ gives $q \, Q(q)^8$ as required. 
   When $m=(1,1,-1)$, the unbroken gauge groups without abelian factors are of the form $\so(n_1)\oplus\so(n_2)\bigoplus_{i=1}^6\spl(2 l_i)$. Here $n_1$ and $n_2$ can be both even or odd, depending on the particular component of the moduli space. Again one can write down the generating functions for the number of states. We saw above that when $m=(1,1,-1)$, only $\omega_3$ has a non-trivial action, so only $e_3$ can be non-trivial (without invoking the $\SL(3,\ZZ)$ symmetry).  As for $m=(1,1,1)$ there are extra correction terms since the large gauge transformations do not always act freely. Furthermore, to compensate for the fact that the total number of branes is $n$ and not $2n$, we let $q\rar q^2$ in the resulting function. We also multiply by an overall factor of $q$. The table below summarise the results:
\bea \label{spgen}
\mbox{\underline{$(e_3,m_3)$}} &\qquad& \mbox{\underline{Generating function}} \non \\
(1,1)     &\qquad& {\ts \frac{1}{8} } q \,Q(q)^8 + {\ts \frac{7}{8} } q \,Q(q^2)^4 \non \\
(-1,1)     &\qquad& {\ts \frac{1}{8} } q \,Q(q)^8 - {\ts \frac{1}{8} }q \,Q(q^2)^4  \non \\
(1,-1)     &\qquad& {\ts \frac{1}{2} }q P(q^2)^2 Q(q^2)^6 + {\ts \frac{1}{2} } q  P(q^4)Q(q^4)^3 \\
(-1,-1)     && {\ts \frac{1}{2} }q P(q^2)^2 Q(q^2)^6 - {\ts \frac{1}{2} } q  P(q^4)Q(q^4)^3 \non \,.
\eea
In summary, the coefficient of $q^{2n+1}$ in the formal Taylor expansion of the expressions in (\ref{spgen}) gives the number of bound states in the $G=\Sp(2n)$ theories with the corresponding $(e_3,m_3)$ values.

$S$-duality now amounts to the statement that the above generating functions should agree under the transformation $(e,m) \mapsto (m,-e)$. Explicitly this requires:
\bea \label{result}
q \prod_{k=1}^{\infty} (1+q^{2n})^8 &\stackrel{?}{=}& {\ts \frac{1}{16} }[\prod_{k=1}^{\infty} (1+q^{2k-1})^8 - \prod_{k=1}^{\infty} (1-q^{2k-1})^8] \non \\
q \prod_{k=1}^{\infty} (1+q^{4n})^4 &\stackrel{?}{=}& {\ts \frac{1}{8} }\prod_{k=1}^{\infty} (1-q^{4k-2})^2[\prod_{k=1}^{\infty} (1+q^{2k-1})^4 - \prod_{k=1}^{\infty} (1-q^{2k-1})^4] \non \\
q \prod_{k=1}^{\infty} (1+q^{2k})^2(1+q^{4n})^4 &\stackrel{?}{=}& {\ts  \frac{1}{8} }[\prod_{k=1}^{\infty} (1+q^{2k-1})^4 - \prod_{k=1}^{\infty} (1-q^{2k-1})^4] \\
q \prod_{k=1}^{\infty} (1+q^{4k})(1+q^{8k})^2 &\stackrel{?}{=}& {\ts \frac{1}{4} }\prod_{k=1}^{\infty} (1-q^{4k-2})[\prod_{k=1}^{\infty} (1+q^{2k-1})^2 - \prod_{k=1}^{\infty} (1-q^{2k-1})^2] \non 
\eea
These complicated expressions can be rewritten in perhaps more familiar form by  recalling the infinite product expansions of the theta functions (theta constants),
\bea \label{thetaexp}
\theta_2(q) &=& 2q^{1/4} \prod_{k=1}^{\infty} (1-q^{2k})(1+q^{2k})^2 \,,\non \\
 \theta_3(q) &=&  \prod_{k=1}^{\infty} (1-q^{2k})(1+q^{2k-1})^2 \,, \\
\theta_4(q) &=& \prod_{k=1}^{\infty} (1-q^{2k})(1-q^{2k-1})^2 \,. \non
\eea
Using these expressions and multiplying the above expressions by some overall factors we see that the first equation in (\ref{result}) can be rewritten as
\be
\theta_2(q)^4 = \theta_3(q)^4 - \theta_4(q)^4 \,,
\ee
which is exactly the theta function version of Jacobi's famous {\it aequatio identica satis abstrusa}!
Similarly the second and third equations can be reformulated as
\be
2\,\theta_2(q^2)^2 = \theta_3(q)^2 - \theta_4(q)^2\,,
\ee
which is a known identity among the theta functions. Finally the last equation in (\ref{result}) can be rewritten as
\be
2\,\theta_2(q^4) = \theta_3(q) - \theta_4(q)\,,
\ee
which again is a known identity among theta functions.

To conclude, we have shown that the spectrum of bound states in the $G=\Spin(2n{+}1)$ and $G=\Sp(2n)$ theories for any $n$ is $S$-duality invariant provided the number of bound states in supersymmetric matrix quantum mechanics agree with the Kac-Smilga conjecture \cite{Kac:1999b}, and vice versa. The appearance of theta functions in the proof was probably accidental, although there may be some string theory calculation that gives the above generating functions more directly and possibly also explains their form and modular properties.

So far in this section, we have only discussed the $S$-duality properties of the genuine bound states (continua of dimension zero in our terminology), but what about the higher-dimensional continua? We can easily generalize the above discussion to incorporate also these states. The general form of the unbroken gauge symmetry for the $G=\SO(2n{+}1)$ and $G=\Sp(2n)$ theories is given by a sum of $\ul(n_i) \simeq \su (n_i) \oplus \ul (1)$ terms and a semi-simple algebra of the same form as before. If we introduce a second variable $y$ to count the number of $\ul(1)$ terms, then since $\su(n_i)$ matrix quantum mechanics has one bound state, the generating function (naively) becomes $\prod_{k=1}^{\infty} (\frac{1}{1-y q^{2k}})$ times the previous generating function. 
This is actually the right result for almost all cases. However, in the $G=\Spin(2n{+}1)$ theories with $e\neq(1,1,1)$ there is a subtlety: 
Recall that one obtains $\ul (1)^l \rightarrow \ul(l)$ gauge enhancement when $l$ of the $(\theta_1^{2 k, 2 k + 1}, \theta_2^{2 k, 2 k + 1}, \theta_3^{2 k, 2 k + 1})$ triples parametrizing the holonomies are equal (modulo signs). 
Now if $l$ is {\it odd}, the action of $\om_i$ on $U_i$ can be undone by shifting  $\tha_i^{ab} \rar \tha_i^{ab} + 2\pi$, so such states have $e=(1,1,1)$. This can be taken into account by letting $q\rar q^2$ in $\prod_{k=1}^{\infty} (\frac{1}{1-y q^{2k}})$. 
Fortunately, this is exactly the replacement that we did in the generating function for the states with $m=(1,1,-1)$ in the $G=\Sp(2n)$ theories. Thus, we conclude that $S$-duality works also for the continua of higher dimensions.

\section*{Acknowledgements}

M.H. is a Research Fellow at the Royal Swedish Academy of Sciences.\\
N.W. is supported by a grant from the Swedish Science Council.

\end{document}